# Reflection on modern methods: Risk Ratio regression - simple concept yet complex computation

**Murthy N Mittinty[1,2]\* John Lynch[1,2,3]**

[1]School of Public Health, The University of Adelaide, Adelaide, Australia.

[2]Robinson Research Institute, University of Adelaide, Adelaide, Australia.

[3]Population Health Sciences, University of Bristol, Bristol, UK.

\*Corresponding author. School of Public Health, University of Adelaide, Adelaide 5005, South Australia. E-mail: murthy.mittinty@adelaide.edu.au

Word count:

Unstructured abstract :150 words

2469 main body without abstract, key words, figure captions and references.

Abstract

The Risk Ratio (RR) is the ratio of the outcome among the exposed to risk of the outcome among the unexposed. This is a simple concept, which makes one wonder why it has not gained the same popularity as the odds ratio. Using logistic regression to estimate the odds ratio is quite common in epidemiology and interpreting the odds ratio as a risk ratio, under the assumption that the outcome is rare, is also common. On one hand, estimating the odds ratio is simple but interpreting it is hard. On the other, estimating the risk ratio is challenging but its interpretation is straightforward. Issues with estimating risk ratio still remains after four decades. These issues include convergence of the algorithm, the choice of regression specification (e.g. log-binomial, Poisson) and many more. Various new computational methods are available that help overcome the issue of convergence and provide doubly robust estimates of RR.

Key Message

- Estimating RR using a simple cross tabulation is easy. However, when it comes to estimating RR using regression, there is no one particular model.
- Use of log-binomial models with continuous covariates may lead to convergence issues.
- Computational methods such as combinatorial expectation maximisation allow convergence of generalised linear models using the binomial family and log link function. However, specification of starting values can be difficult.
- The binary regression method which allows direct modelling of risk ratios may be a better choice.

**Introduction**

Relative risk is a common term used in epidemiology to refer to risk and rate ratios.[1] The concept of risk ratio (RR) when introduced first to students is taught using a simple $2 \times 2$ table and a hand calculator. The $2 \times 2$ table is created using a simple cross tabulation of a binary exposure and a binary outcome. Using the information from this cross tabulation, RR is estimated as the ratio of risk of the outcome among the exposed versus the risk of the outcome among the unexposed. For example, let's say the outcome is low birthweight (Yes = 1 or No = 0) and the exposure is maternal smoking during pregnancy (Yes = 1 and No = 0). Risk ratio, in this example, is the ratio of the proportion of low birthweight children among smokers to the proportion of low birthweight children among non-smokers. In this form it is simple and easy to calculate.

Let's consider adjusting for one confounder like gender which is binary; in this case, RR can be estimated within the stratum of gender. Now suppose there is a long list of confounders which includes age, education, income, pregnancy related factors and others. To estimate RR in this case, one may need to use regression. Use of regression methods for estimating RR gained popularity when they became available in regular commercial and non-commercial statistical software. Even with this availability, it is still not free from problems which has concerned researchers since the 1980s.[2] Other methods such as logistic regression gained immense popularity and have become essential tools in epidemiology due to the computational ease, and as the odds ratio (OR) can approximate the RR in the case of rare events. Evidence suggests that logistic regression is used to estimate the OR but is commonly interpreted as RR.[3] However, OR overestimates RR, whenever RR is greater than 1, and hence should not be interpreted as RR.[4,5,6]

If logistic regression is used to estimate RR under the rare disease assumption, then one must note that this assumes that the conditional probability of having an outcome given the unexposed state (baseline prevalence, $p(Y = 1|X = 0) = p_0$) approaches zero (as shown in web supplement S1). Moreover, as suggested by the reviewer, relation between OR and RR can be derived as shown in S1, using this derivation if we assume $RR^{max} = 10$ (*upper bound*) and $p_0 = 0.001$ we have $\frac{OR}{RR} \leq 1.01$. Thus, if the $RR^{max}$ is less than or equal to 10 and the baseline prevalence is 1 in 100- then the relative error OR/RR is 1%. With a prevalence of 1 in 10000 it is 0.1%, when the prevalence is very small but not zero, the approximation errors are small enough to be practically negligible. We assume the RR >1 but less than some maximum plausible value $RR^{max} > 1$.

Alternatively, let's examine this using a simple $2 \times 2$ table with four cells. Let these cells be labelled as *a, b, c* and *d*, where '*a*' is the count when the outcome is 1 and the exposure is 1, '*b*' is the count when the exposure is 1 and outcome is 0, '*c*' is the count when the outcome is 1 and the exposure is 0 and '*d*' is when both outcome and exposure are 0. Now to estimate RR, we use the formula $\frac{\frac{a}{a+b}}{\frac{c}{c+d}}$. If we rearrange the terms, we estimate the RR as $\frac{a*(c+d)}{c*(a+b)} = \frac{ac+ad}{ca+bc}$, whereas the OR is estimated as $\frac{ad}{bc}$. Again, from these formulae, one may note that RR does not equal (or even approximate) OR without some assumptions. One common assumption can be that the outcome is rare in both the groups of the exposure (if exposure is the only variable, else, the outcome of interest must be rare for all the levels of the covariates). Furthermore, let's put some numbers instead of *a, b, c, d*, say a = 1, b = 5, c = 1 and d = 11. In this case, the estimate of RR using the above formula equates to 12/6 = 2 which is the ratio of the marginal totals of the exposure when $X = 1 (a + b)$ and $X = 0 (c + d)$. Now, if we estimate the OR (= 2.2), as shown in the supplement S2 the OR equates to the ratio of not having the outcome when the exposure is absent versus not having the outcome when exposure is present. In this example, equating OR and RR may not be appropriate as they are estimating two different things. Moreover, both OR and RR are not estimating the risk of disease whenever the counts *a* and *c* are equal in a $2 \times 2$ table or a stratified $2 \times 2$ table. In summary, if the study outcome is common, interpretation of OR as an approximation to RR becomes unreliable.

Odds ratios may still be of interest because they are symmetric, in the sense that the odds of having an outcome is the inverse of odds of not having the outcome (mathematically this might be interesting but practically, when the outcome is defined as death or survival this property might not seem desirable), and when the covariate set is large it may be a preferred choice.[7] Moreover, in some case-control studies when studies use cumulative incidence sampling OR maybe valuable.[8] On the other hand, RR is not symmetric (with respect to relabelling of the outcome *Y*) but the size of the RR will not change if adjustment is made for a variable that is not a confounder. This is referred to as collapsibility. Collapsibility property implies that the risk ratio can be expressed as the ratio change in average risk due to exposure among the exposed.[7,9 10,11,12] It is for this reason RR, and for its ease of interpretation, maybe a preferred parameter of interest over OR.

Several methods have been proposed to estimate the RR. These include the Stratified Mantel-Haenszel method[4], Cox regression[13,3], adjustment to OR[14] (even though this method was later noted to be biased[15]), Generalised Linear Models (GLM) with family binomial and link log, referred to as log-binomial.[16,17] However the log-binomial method has the issue of convergence[2,3,16] in STATA, R, SAS, Splus or any other software.

To overcome this issue, methods such as the COPY method[2,13,16], modified Poisson[18], marginal standardization[16,17,19], binary regression models[9], quasi-likelihood Poisson method[20] constrained optimization[21] and non-linear least squares[3] have been proposed. Some of the software available include libraries such as *logbin* (log binomial models)[21, 22, 23] and *brm* (binary regression model) in R[9].

This raises the question: if RR is a simple concept, why don't regression methods, using MLE with standard Fisher scoring matrix, converge when estimating RR? Are there different computational methods? How should the results be presented? We provide some answers to these questions.

**Why are there different methods?**

Let *Y* be the binary outcome of interest, *X* the binary exposure and *C* be the vector of confounders. *Y* is 1, representing the occurrence of an event and 0 represents the non-occurrence. Similarly, when the exposure equals 1, we say the individual is exposed/treated and 0 indicates those non-exposed/untreated. Confounders can be continuous, categorical or binary variables (examples include age, levels of gender). The success probability in RR regression is modelled as:

$$\log(P[Y_i = 1|D_i = (X_i, C_i)]) = \beta_0 + \beta_1 X_i + \beta_2 C_{1i} + \cdots + \beta_p C_{pi} = (\boldsymbol{\beta D_i})$$

Denote $P[Y_i = 1|D_i] = p_i, \forall\ i = 1,2,\ldots,n$, as the probability of having an outcome for *n* individuals in the data (*D*). The above equation can be rearranged in a matrix form and written as

$$\log(p_i) = \boldsymbol{\beta D_i}$$

Using the relation between natural logarithms and exponentials, the above equation can be expressed as, $p_i = e^{\boldsymbol{\beta D_i}}$. Here, the parameter $\boldsymbol{\beta}$ is unknown and this vector needs to be estimated. To estimate the unknown parameter, we will use the Bernoulli likelihood function which is given by

$$L(\beta) = \Pi_{i=1}^{n} p_i^{Y_i}(1-p_i)^{(1-Y_i)}.$$

Various methods to estimate/fit the model include the maximum likelihood estimating (MLE) equation for $\beta$, obtained by taking the derivative of the logarithm of the above likelihood function ($L(\beta)$) and equating it to zero. Mathematically this is simplified as (for complete derivation, see web supplement S3)

$$S(\beta) = \frac{\partial \log L(\beta)}{\partial \beta_j} = \sum_{i=1}^{n} \frac{(Y_i - p_i)}{p_i(1-p_i)} d_i p_i = \sum_{i=1}^{n} \frac{d_i(Y_i - p_i)}{(1-p_i)} \quad (1)$$

where $d_i$ is a realisation of vector $D_i$. This is an *asymptotically* efficient estimate, when the probability of success ($p_i$) is less than 1 then the MLE exist, and it is unique. However, when $p_i \approx 1$ then the estimating function will be dominated by observation $i$, and convergence issues persist. When the MLE does not converge some software uses constrained optimisation techniques as a default solution and thus attains convergence.[3]

In standard software MLE is computed using methods like the Newton Raphson method, iteratively reweighted least squares (IRWLS) and Fisher scoring.[21,22,23] However these computational methods have issues when the probability approaches 1 in log-binomial models.[2,3,23]

Alternatively, the modified Poisson regression method[15] has been proposed for estimating $\beta$ and has gained attention. The MLE for the Poisson regression is given by

$$S_{Poisson}(\beta) = \sum_{i=1}^{n} d_i(Y_i - p_i)$$

As seen from above notation, the Poisson regression does not suffer from the issue of convergence as there is no denominator which may approach zero. Now the question is: How does one get rid of the denominator in Equation 1? To understand this, it is important to take a step back and revisit the concept of Maclaurin series. Using the Maclaurin series, $\frac{1}{1-p_i}$, in Equation 1 can be expressed as

$$\frac{1}{1-p_i} = 1 + p_i + p_i^2 + p_i^3 + \cdots = \sum_{m=0}^{M} p_i^m. \qquad (2)$$

Replace $\frac{1}{1-p_i}$ in Equation 1 as the weight, $w(p_i, M)$, then this can be re-expressed as:

$$S(\beta) = \sum_{i=1}^{n} d_i(Y_i - p_i) w(p_i, M),$$

when $M = 0$ in Equation 2, then the weight, $w(p_i\ M) = 1$. Hence, the RR estimated using a Poisson regression can be viewed as Maclaurin series truncated at $M = 0$. However, with binary outcomes not all combinations of parameters lead to fitted means that are between zero and one. It allows for higher values of $M$ to be used. In 2014, Fitzmaurice et al.[24] proposed a method that uses $M = 20, 30, 40$ and $60$, also known as the "*almost efficient estimation of RR*". Thus, all variants of the weighted regression, including when $M$ equals 0, will only estimate RR almost efficiently, but not completely efficiently.

If using Poisson and interpreting results from this regression, then one must specify it as a truncated ($M = 0$) Maclaurin series. If using higher terms, as done by Fitzmaurice et al[24], then we must say that exactly ($M = 60$). When Poisson regression is applied to binomial data, the standard error for the estimated RR will be overestimated.[15] To overcome this issue one can use the sandwich estimation procedure to compute the robust error variance.[19,26] However, when the sample sizes are small Poisson models do not work well because the sandwich estimators tend to underestimate the true standard errors.[27,28] Furthermore, one of the weakness of sandwich estimators is that their variance can be less efficient than the variance estimated from a parametric model.[28] This weakness then impacts the coverage probability, the probability that a confidence interval covers the true RR.[29] Moreover, the predicted probabilities using Poisson regression can lie outside of the range [0,1].[6,27] This happens because RR is variation dependent on the baseline probability ($p(Y = 1|X = 0)$). For example, if $RR = 2$, then $p(Y = 1|X = 1) = 2 * p(Y = 1|X = 0)$, this indicates that $p(Y = 1|X = 0) \leq 0.5$. This is a restricted domain over which the quantities $(RR, p(Y = 1|X = 0))$ need to be compatible with a valid probability distribution. As can be observed from this example, it is not only for the Poisson regression, even in log-binomial models, with considerable number of covariates, finding MLE can be a problem as the parameter space is constrained and the log likelihood (Equation 1) needs to be maximised using constrained optimization.[3]

The next questions that naturally arise are 1) how to achieve convergence in log-binomial models; and 2) Presenting results from multiple methods.

**How to achieve convergence in log-binomial models?**

With the log binomial fitting procedure one can start by increasing the maximum number of iterations along with specification of starting values. The starting values can be set to the mean observed proportion for the intercept and rest all parameters can be set to zero. However, if the standard default procedures (IRWLS algorithm, Newton-Raphson or Fisher Scoring computational methods) are used in estimating the MLE of the log-binomial then they may not converge, in such situations computational methods like combinatorial expectation maximisation (CEM), adaptive barrier method, Parabolic expectation maximisation (PEM) /quasi Newton methods may be used through packages such as *logbin* in R.[21,22,23] Use of these latter computational methods may allow convergence if the starting values are specified. [17] Coming up with a proper set of starting values can be tricky. If the starting values are appropriate, then there is a chance of attaining convergence else not. For CEM, if the covariate set is large then again there will be no convergence. With alternative methods convergence still persists because of the constrained optimisation.[9]

Alternatively, one can use the BRM approach which overcomes the variance dependence and constrained optimization.[9] The BRM allows direct modelling of RR.[9] BRM uses two different regressions 1) an outcome regression and 2) a propensity score regression of the exposure on the baseline covariates. Furthermore, the outcome regression uses two different models: a) target model for estimating RR directly and b) a nuisance model for the log odds product. Use of the log odds product allows estimation of RR either using an unconstrained MLE or semiparametric g-estimation methods. If the target model is correctly specified and either the log odds product model or the propensity score model is correctly specified BRM yields a robust estimate.[9] Similar to *glm* methods BRM also requires specification of starting values and may converge to local maxima.

**Presenting results**

For this demonstration, we use data from the National Health and Nutrition Examination Survey Follow-up Study to estimate the RR in the covariate adjusted associational sense. This data is available as accompanying data to the book by Hernan and Robin.[30] We are primarily interested in the association between quitting smoking (Yes/No) and a dichotomised weight change (above and below median weight) between 1971 and 1982. Code that is required to run all analysis and reproduce the results presented here is available in the supplement S4. In our analysis we adjusted for sex, age, race, income, marital status, education, asthma, and bronchitis. All analysis was conducted in R version 3.6.3 and Stata 15.1.

Results from these methods are presented in Table 1.

Table 1: Risk Ratio estimates for the association between Quitting smoking and greater than Median Weight Gain Among, 1629, Men and Women in the National Health and Nutrition Examination Survey Epidemiologic Follow-up Data between 1971-1982.

| Method[a] | Estimation | Computation | Risk Ratio (95% CI) |
|---|---|---|---|
| Mantel Haenszel (Combined) | | | 1.28 (1.17,1.41) |
| GLM[b] (Family=binomial, link=log) | MLE[d] | IRLS[f] | NC[j] |
| GLM (with defined starting values) (Family=binomial, link=log) | MLE | IRLS | 1.29 (1.18,1.43) |
| GLM (Se[c] = Sandwich) (Family=Poisson, Link=log) | MLE | IRLS | 1.34 (1.22,1.48) |
| Binary Regression model | MLE | | 1.36 (0.98,1.73) |
| Binary Regression model | DR[e] | | 1.39 (0.86,1.91) |
| Logbinomial | MLE | EM[g] (CEM[h]) | 1.32 (1.20,1.45) |
| Logbinomial | MLE | AB[i] | 1.32 (1.20,1.45) |

[a]All models, except Mantel Haenszel, adjusted for age, sex, race, income, marital status, education, asthma, and bronchitis.
[b]GLM: Generalised Linear Model
[c]Se: Standard error
[d]MLE: Maximum Likelihood Estimation
[e]DR: Doubly Robust Estimation
[f]IRLS: Iterative Reweighted Least Squares
[g]EM: Expectation Maximization
[h]CEM: Combinatorial Expectation Maximization
[i]AB: Adaptive Barrier.
[j]NC: Non Convergence

**Conclusion**

In general, RR can be estimated using a hand calculator if presented as a simple 2x2 table. A common problem when using regression to estimate RR is lack of convergence of the MLE. With the provision of additional computational methods such as CEM, adaptive barrier, or other methods as alternatives to standard methods such as IRWLS, we may overcome the issues of convergence in the log-binomial model if proper starting values are specified and the covariate set is small. When the covariate set is large then one can use the BRM method which allows direct modelling of RR. Use of BRM for estimating RR may produce wider (conservative) confidence intervals for the RR. However, further research directly comparing the RR estimates between BRM and *glm* methods need to be conducted.

# Appendix

## S1

Let's take a look at the error of approximating the OR to RR in terms of probabilities in a slightly different way. The RR is usually computed as $RR = p(Y = 1|X = 1)/p(Y = 1|X = 0)$, where Y=1 is having outcome and X=1 is referred to as having the exposure. P(Y=1|X=1) is the conditional probability of having an outcome given that the respondent/individual has been exposed. RR is the ratio of the outcome among the exposed to risk of the outcome among the unexposed. Similarly, the odds ratio (OR) is expressed as $OR = \frac{(p(Y = 1|X = 1)p(Y = 0|X = 0))}{(p(Y = 1|X = 0)p(Y = 0|X = 1))}$. Now, the relation between OR and RR can be rederive it as

$$OR = RR\left(\frac{1 - p(Y = 1|X = 0)}{1 - p(Y = 1|X = 0)RR}\right)$$

Thus, the relative approximation error is then given as

$$OR/RR = \left(\frac{1 - p(Y = 1|X = 0)}{1 - p(Y = 1|X = 0)RR}\right)$$

**Case where $RR < 1$**

When RR < 1 and for a fixed baseline prevalence, $P(Y = 1|X = 0)$, say $p_0$, the relative approximation error is $1 > \frac{OR}{RR} > 1 - p_0$, with the largest discrepancy (corresponding to the smallest value of OR/RR as $RR \rightarrow 0$).

**Case where $RR > 1$**

The case when $RR > 1$, for fixed $p_0$ the relative error increases without limit as the $RR \rightarrow \frac{1}{p_0}$.

Suppose $RR^{max} = 10$ and $p_0 = 0.001$ then using $\left(\frac{1-p(Y=1|X=0)}{1-p(Y = 1|X = 0)RR^{max}}\right)$ we get the value as $\frac{1-0.001}{1-0.001*10} = \frac{0.999}{0.99} = 1.009 = 1.01$.

Say if we have $RR^{max} = 5$ and the prevalence is 0.0001 then we have $\frac{1-0.0001}{1-0.0001*5} = \frac{0.9999}{0.9995} = 1.0004$. However, if the prevalence is 0.1 and $RR^{max}$ is 5 then we have $\frac{1-0.1}{1-0.1*5} = \frac{0.9}{0.5} = 1.8$. Relative error is 80% here. That is in the worst-case scenario the OR is 1.8 times as large as the RR. Here the question can be what is $RR^{max}$ and how does one get this value? Suppose that we know the value of the OR, for example from case-control data, but we are interested in the RR. Further suppose that we know the baseline prevalence ($p_0$), we believe that if the alternative is true then the $RR > 1$, and that we have an idea about the maximum possible size of the RR, which we will call $RR^{max}$; This could be based on experience with this disease or others with similar etiology. In other words, $RR^{max}$ is a worst-case scenario in terms of the approximation error. If in fact the RR is much closer to 1 than $RR^{max}$ then the approximation error in using the OR in place of the RR will be small.

## S2

Let *Y* be the outcome and *X* be the exposure, assume that the exposure is a binary variable and the outcome is also a binary variable. The simple cross tabulation of these variables yields a 2x2 table as

| Outcome | Exposure | | Total |
|---|---|---|---|
| | $X = 1$ | $X = 0$ | |
| $Y = 1$ | $a = 1$ | $c = 1$ | $2 = n_1$ |
| $Y = 0$ | $b = 5$ | $d = 11$ | $16 = n_2$ |
| Total | $a + b = 6$ | $c + d = 12$ | $N = 18$ |

$RR = \frac{\frac{a}{a+b}}{\frac{c}{c+d}} = \frac{\frac{1}{6}}{\frac{1}{12}} = \frac{12}{6}$, from the above table one may note that these numbers are nothing but the frequencies of being exposed and unexposed. Now if these counts (*a, b, c,* and *d*) are divided by total sample $N = 18$ then the above RR is expressed in terms of probability as $\frac{\frac{12}{18}}{\frac{6}{18}} = \frac{p(x=0)}{p(x=1)}$. This shows that whenever the counts *a* and *c* are equal or also rare RR is estimating the ratio of the exposure distribution.

$$OR = \left(\frac{ad}{bc}\right) = \frac{1 * 11}{1 * 5} = \frac{11}{5} = 2.2$$

Let's look at the formula in terms of the probability. Count *a* refers to the joint distribution of having exposure and also the outcome, denote this as $J(y = 1, x = 1)$. When we divide *a* by the total number *N* we can call the fraction as the joint probability denoted as $p(y = 1, x = 1)$. Similarly, *b/N* is $p(y = 1, x = 0)$; $\frac{c}{N} = p(y = 1, x = 0)$ and *d/N* is $p(y = 0, x = 0)$. We now are ready to estimate OR in terms of probabilities from the above 2x2 table as.

$$\begin{aligned} OR &= \frac{p(y = 1, x = 1) \times p(y = 0, x = 0)}{p(y = 1, x = 0) \times p(y = 0, x = 1)} \\ &= \frac{p(x = 1)p(y = 1|x = 1) \times p(x = 0)p(y = 0|x = 0)}{p(x = 0)p(y = 1|x = 0) \times p(y = 0|x = 1)p(x = 1)} \\ &= RR \left(\frac{p(y = 0|x = 0)}{p(y = 0|x = 1)}\right) = \frac{12/18}{6/18} * \left(\frac{11/18}{12/18} * \frac{6/18}{5/18}\right) = \frac{11/18}{5/18} = 2.2 \end{aligned}$$

The final estimate, in this example, OR is equating to the ratio of the joint distribution of the absence of outcome when exposed and unexposed. These two estimates, in this example, as one can see are estimating two different things and hence must not be approximated. This is because they are answering two different questions 1) What is the ratio of having an exposure to not having an exposure (RR case) and 2) What is the joint probability of not having an outcome when not exposed to the joint probability of not having an outcome when exposed (OR case). Compared to 1) What is the ratio of the probability of having the disease if exposed to the probability of having the disease if not exposed and 2) What is the ratio of the odds of having the disease when exposed to the odds of having the disease when not exposed. Hence one need to be careful when interpreting or approximating these values.

## S 3
**Derivation of the log likelihood**

Maximum likelihood estimation of the log-binomial model is

$$l(\beta) = \sum_{i=1}^{n} y_i \log(p_i) + \sum_{i=1}^{n}(1-y_i)\log(1-p_i)$$

$$\frac{\partial l(\beta)}{\partial \beta_j} = \sum_{i=1}^{n} \frac{y_i}{p_i}\frac{\partial p_i}{d\beta_j} - \sum_{i=1}^{n}\frac{1-y_i}{1-p_i}\frac{\partial p_i}{d\beta_j}$$

$$= \sum_{i=1}^{n}\left(\frac{y_i}{p_i} - \frac{1-y_i}{1-p_i}\right)\frac{\partial p_i}{\partial \beta_j} = \sum_{i=1}^{n}\frac{(y_i - p_i)}{p_i(1-p_i)}\frac{\partial p_i}{\partial \beta_j}$$

where $p_i = e^{x_i \beta_j}$

**S 4**

**Risk Ratio (RR) estimation in Stata and R**

The following section will provide details of estimating RR using the listed approaches, which have been classified into three groups: 1) Mantel-Haenszel stratified method for a small set of covariates (in this same group, we also have listed the log-binomial models and variants of log-binomial); 2) logbin regression with various computational methods; and 3) the Non-GLM, doubly robust method.

**#Reading data into R and making necessary manipulation to attain results we presented in Table 1 of the manuscript**

*Nd<- url("https://cdn1.sph.harvard.edu/wp-content/uploads/sites/1268/1268/20/nhefs.csv")*

*nd<-read.csv(Nd)*

#data manipulations implemented to make income, marital status and outcome as binary variables

*nd$incomeb<-ifelse(nd$income>15,1,0) #income binary*

*nd$maritalb<-ifelse(nd$marital>2,1,0) #marital status binary*

*nd$wtb<-ifelse(nd$wt82_71>median(nd$wt82_71,na.rm=TRUE),1,0) #weight binary*

*nd<subset(nd,select=c(qsmk,wtb,exercise,sex,age,race,incomeb,maritalb,school,asthma,bronch))*

*factor_names <- c("exercise","incomeb","maritalb","sex","race","asthma","bronch")*

*nd[,factor_names] <- lapply(nd[,factor_names] , factor)*

*formulaVars <- paste(names(nd)[-c(2)],collapse = "+")*

*modelForm <- as.formula(paste0("wtb ~", formulaVars))*

*modelForm*

**Copy the data into Stata to do the Mantel Haenszel and GLM models**

**Mantel Haenszel, log-binomial and variants**
*Method: Stratified Mantel Haenszel method*
Useful when there is one confounder apart from the exposure. Stata code is

*Clear*
*Use nd*
*cs wtb qsmk, by(sex) pool*

**Going back to R program for conducting the GLM, logbin and BRM methods**

**Log-Binomial**
*Method: Log binomial model*
Log-binomial is a natural choice for developing log linear model and directly estimate RR. When GLM methods have convergence issues then. RR can be estimated using the following method
**Standard form if there is no convergence issue**

   *bin_id <- glm(modelForm,data=nd,family = binomial("log"))*
   *bin_id*

If the above command has convergence issues then one can force convergence using the following code

   *bin_id <- glm(modelForm,data=nd,family =*
   *binomial("log"),start=c(log(846/1629),rep(0,11)))*
   *bin_id*

**S 4**

**Alternate methods**
*Method: Logbin regression in R.*
*#installing packages*
   *install.packages("logbin")*
   *install.packages("sandwich")*
   *library(sandwich)*
   *library(logbin)*

#logbin regression with adaptive barrier (constrained optimisation) computational method

   *start.p<-c(log(846/1629),cf)*
   *fit.logbin <- logbin(formula(bin_id), data = nd,*
            *start = start.p, trace = 1,method="ab")*

#Extracting starting values from a Poisson model (we used these in the model)

   *modelRR <- glm(modelForm,data=nd,family = poisson("log"))*

   *cf<-modelRR$coefficients*

   *cf<-cf[-1]*

#logbin regression with the Expectation maximization algorithm

```
start.p<-c(log(846/1629),cf)
fit.logbin <- logbin(formula(bin_id), data = nd,
            start = start.p, trace = 1,method="ab")
fit.logbin.em <- update(fit.logbin, method = "em")
```
# Speed up convergence by using acceleration methods
```
fit.logbin.em.acc <- update(fit.logbin.em, accelerate = "squarem")
fit.logbin.em.acc
```

**S 4**

*Method: binary regression model in R*
```
install.packages("brm")
library(brm)
y<-nd$wtb
x<-nd$qsmk
v<-nd[,-c(1,2)]
int<-rep(1,nrow(v))
v<-cbind(int=int,v)
v<-as.matrix(v)
fit.mle=brm(y,x,v,v,'RR','MLE',v,TRUE)
fit.mle

fit.drw = brm(y,x,v,v,'RR','DR',v,TRUE)
fit.dru = brm(y,x,v,v,'RR','DR',v,FALSE)
```

***The parameter RR its mean and standard deviation can be computed using the following code***

```
mean(fit.drw$param.est)
mean(fit.dru$param.est)
sd(fit.drw$param.est)
sd(fit.dru$param.est)
```